\documentclass[11pt,twoside]{article}
\usepackage{asp2010}

\resetcounters

\newcommand{\msol}{\mathcal{M}_\odot}
\newcommand{\muu}{mag arcsec$^{-2}$}

\newcommand{\sfr}{\dot{\mathcal{M}}_\ast}

\newcommand{\Ssfr}{\dot{\Sigma}_\ast}
\newcommand{\Ssfre}{\dot{\Sigma}_{e,\ast}}
\newcommand{\Sge}{\Sigma_{e,g}}
\newcommand{\Sse}{\Sigma_{e,\ast}}

\newcommand{\sdmh}{\Sigma_{\rm H_2}}
\newcommand{\sdhi}{\Sigma_{\mbox{\rm \footnotesize H{\scshape i}}}}

\newcommand{\qrw}{Q_{\rm RW}}
\newcommand{\qrwmin}{\qrw^{\rm min}}

\begin{document}

\title{The Stability of Galaxy Disks}

\author{ Kyle B.~Westfall,$^1$ David R.~Andersen,$^2$ Matthew A.~Bershady,$^3$
Thomas P.~K.~Martinsson,$^4$ Robert A.~Swaters,$^5$ \& Marc A.~W.~Verheijen$^1$
\affil{$^1$Kapteyn Astronomical Institute, University of Groningen}
\affil{$^2$NRC Herzberg Institute of Astrophysics}
\affil{$^3$Department of Astronomy, University of Wisconsin-Madison}
\affil{$^4$Leiden Observatory, Leiden University}
\affil{$^5$National Optical Astronomy Observatory}
}

\begin{abstract}
We calculate the stellar surface mass density ($\Sigma_\ast$) and two-component
(gas$+$stars) disk stability ($\qrw$) for 25 late-type galaxies from the
DiskMass Survey.  These calculations are based on fits of a dynamical model to
our ionized-gas and stellar kinematic data performed using a Markov Chain Monte
Carlo sampling of the Bayesian posterior.  Marginalizing over all galaxies, we
find a median value of $\qrw=2.0\pm0.9$ at 1.5 scale lengths.  We also find that
$\qrw$ is anti-correlated with the star-formation rate surface density
($\Ssfr$), which can be predicted using a closed set of empirical scaling
relations.  Finally, we find that the star-formation efficiency
($\Ssfr/\Sigma_g$) is correlated with $\Sigma_\ast$ and weakly anti-correlated
with $\qrw$.  The former is consistent with an equilibrium prediction of
$\Ssfr/\Sigma_g \propto \Sigma_\ast^{1/2}$.  Despite its order-of-magnitude
range, we find no correlation of $\Ssfr/\Sigma_g\Sigma_\ast^{1/2}$ with any
other physical quantity derived by our study.
\end{abstract}

\noindent{\bf Motivation:} Studies of the star-formation law in disk galaxies
have largely focused on assessments of the gaseous component
\citep[e.g.,][]{1998ApJ...498..541K}.  This approach is understandable given
that it is the gas from which stars are formed.  However, the stellar component
is also relevant; for example, it is often the dominant contributor to the
gravitational potential in the disk plane.  Indeed, a relation between the
star-formation efficiency (SFE; the star-formation rate per unit gas mass) and
the stellar surface mass density ($\Sigma_\ast$) has been found empirically
\citep{2011ApJ...733...87S} and is expected theoretically
\citep{2010ApJ...721..975O}.

The theory presented by \citet{2010ApJ...721..975O} is derived assuming an
equilibrium of the diffuse and self-gravitating gas with respect to its thermal
properties and the pressure balance within the vertical gravitational field of
the disk.  Depending on the relevant timescales, this equilibrium may not be
reached in galaxies that exhibit modal potential perturbations (e.g. spiral
arms).  In so far as the two-component disk stability ($\qrw$; see below)
quantifies the susceptibility of a disk to such perturbations, it is therefore
interesting to test the validity of the equilibrium prediction in disks of
different $\qrw$.

The primary systematic uncertainties in most extant calculations of
$\Sigma_\ast$, and $\qrw$, are incurred via the use of
stellar-population-synthesis models \citep[e.g.][]{2008AJ....136.2782L}.  In
contrast, the high-resolution stellar kinematic data from the DiskMass Survey
\citep{2010ApJ...716..198B} allow for a {\it dynamical} calculation of
$\Sigma_\ast$, which is not subject to the same systematic errors.  Therefore,
we use these data to investigate the correlation of $\Ssfr$ and SFE with
$\Sigma_\ast$ and $\qrw$.

\begin{figure}[!ht]
%
\plotfiddle{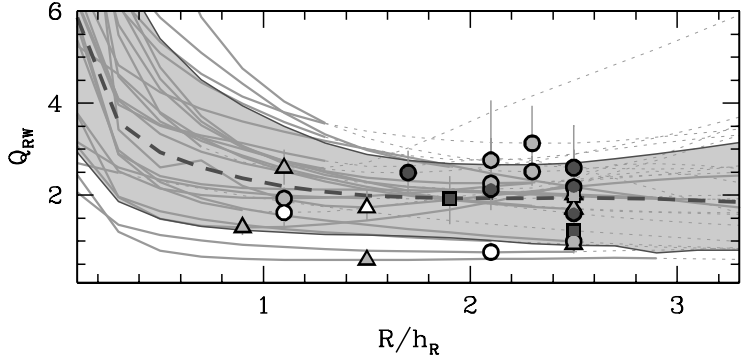}{4cm}{0}{85}{85}{-250}{-308}
\caption{
The two-component disk stability, $\qrw$, as a function of $R/h_R$ from the
dynamical model of each galaxy.  The profile for each galaxy transitions from
the solid- to dotted-gray lines at the radius when the model is no longer
directly constrained by our LOS stellar velocity dispersions, only by the
rotation curves.  The minimum $\qrw$ within 2.5 $h_R$, $\qrwmin$, is marked for
each galaxy: light-gray is used for ``bc'' and ``c'' type spirals; white and
dark-gray are used for earlier and later Hubble types, respectively.  Circles,
triangles, and squares are for unbarred (S), weakly barred (SAB), and barred
(SB) galaxies, respectively.  The dark-gray dashed line is the median $\qrw$
from the marginalized distributions at each $R/h_R$, and the light-gray region
is the 68\% confidence interval.
}
\label{fig:fig1}
\end{figure}

\medskip\noindent{\bf Measurements:} Detailed discussions of our dynamical
assumptions can be found in papers from the DiskMass Survey series
\citep[e.g.,][]{2010ApJ...716..234B, 2011ApJ...742...18W, 2013A&A...557A.131M}.
Briefly, we calculate the dynamical surface mass density, $\Sigma_{\rm dyn}
\propto \sigma_z^2/h_z$, assuming a parallel-plane disk with an exponential
vertical density profile \citep{1988A&A...192..117V}.  To calculate the scale
height ($h_z$), we use a scaling relation between the disk oblateness
($h_R/h_z$) and its scale length ($h_R$) based on observations of edge-on
spirals \citep{2010ApJ...716..234B}.  To obtain the vertical velocity dispersion
($\sigma_z$), we fit a dynamical model to our line-of-sight (LOS) kinematic data
that yields the shape of the stellar velocity ellipsoid.  We obtain
$\Sigma_\ast$ by subtracting the gas mass surface density, $\Sigma_g =
1.4(\sdmh+\sdhi)$, from $\Sigma_{\rm dyn}$.  Finally, we calculate the
\citet{1964ApJ...139.1217T} stability criterion, $Q_i\propto
\kappa\sigma_{R,i}/\Sigma_i$, for the gas and stars individually based on the
results of the dynamical model and combine them into a two-component stability
($\qrw$) following \citet{2011MNRAS.416.1191R}; $\kappa$ is the epicyclic
frequency, and the cold-gas velocity dispersion is assumed to be isotropic and
half of the ionized-gas dispersion.  The assumptions made by the dynamical model
are very similar to those from \citet{2011ApJ...742...18W}, but the methodology
follows Bayesian statistics (see Westfall et al., {\it in prep}).

Figure \ref{fig:fig1} shows $\qrw(R)$ for each galaxy individually and when
marginalized over all galaxies.  The marginalized $\qrw$ is large toward the
center ($\kappa$ is largest in the rising part of the rotation curve) and then
asymptotes to a nearly constant value at $R>1h_R$; the median of the
marginalized probability distribution is $\qrw=2.0\pm0.9$ at $R=1.5 h_R$.

\begin{figure}[!ht]
\plotone{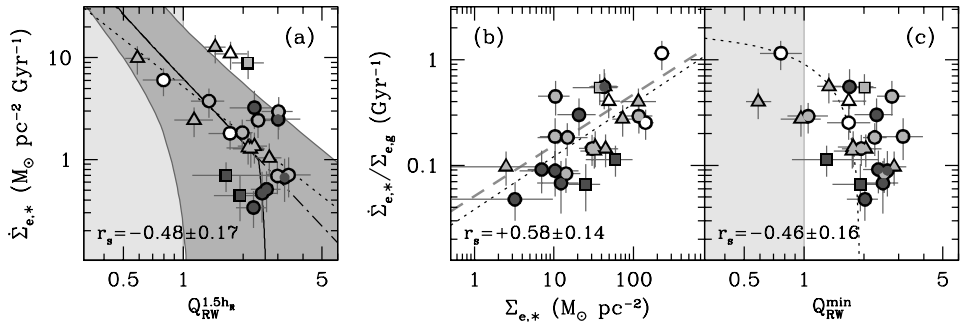}
\caption{
{\it (a)} Measurements of $\Ssfre$ and $\qrw^{1.5h_R}$ for our galaxies with the
nominal disk instability region in light-gray.  Colors and symbol types are the
same as in Figure \ref{fig:fig1}.  The solid black line is the predicted
correlation based on empirical scaling relations when using the average
properties of our sample; the dark-gray region encompasses results found when
using the parameters appropriate for each galaxy.  The dotted line is the
expectation from \citet{2006ApJ...639..879L} with an optimal normalization for
our data. {\it (b)} SFE versus $\Sse$ and {\it (c)} SFE versus $\qrwmin$.  See
text for more description.
}
\label{fig:fig2}
\end{figure}

\medskip \noindent{\bf Anti-correlation Between Stability and Star-formation
Activity:} We calculate $\Ssfre = \sfr/\pi R_{25}^2$ using star-formation rates
($\sfr$) based on 21-cm radio-continuum measurements and the calibrations from
\citet{2001ApJ...554..803Y}, where $R_{25}$ is the radius of the $\mu_B = 25$
\muu\ surface-brightness isophote.  The results are compared with $\qrw$ at
$1.5h_R$ ($\qrw^{1.5h_R}$) in Figure \ref{fig:fig2}a.  The Spearman
rank-correlation coefficient, $r_s$, demonstrates that the two quantities are
anti-correlated; however, the correlation is only roughly three times its
measurement error, as estimated using bootstrap simulations.  The
anti-correlation strengthens to $r_s = -0.53\pm0.14$ if we instead consider the
minimum $\qrw$ within $R\leq2.5h_R$ ($\qrwmin$; see Figure \ref{fig:fig1}).  The
anti-correlation between $\Ssfre$ and $\qrw$ can be predicted based on empirical
scaling relations.

The details of the scaling-relation calculation will be presented by Westfall et
al.\ ({\it in prep}).  In short, we define a set of auxiliary parameters ---
cold-gas dispersion, $\sigma_g$; central disk surface brightness in $K$-band,
$\mu_{0,K}$; $K$-band mass-to-light ratio, $\Upsilon_K$; $h_R$; $R_{25}$; and
$\alpha=\sigma_z/\sigma_R$ --- that, for a given $\sfr$, can be used to
determine the input quantities required in the calculation of $\qrw(R)$ ---
$\kappa$, $\sigma_g$, $\sigma_R$, $\Sigma_g$, and $\Sigma_\ast$.  For the black
line in Figure \ref{fig:fig2}a, we use the Kennicutt-Schmidt law
\citep{1998ApJ...498..541K} to determine the average $\Sigma_g$ within $R_{25}$
for a given $\Ssfre$ and distribute that gas according to the $\Sigma_g (R)$
profile from \citet{2012ApJ...756..183B}.  We assume a hyperbolic tangent form
for the circular speed --- used to get $\kappa$ \citep{2008gady.book.....B} ---
with parameters that follow the scaling relations with the light profile from
\citet{2013ApJ...768...41A}, and yields a ratio with a disk-only rotation curve
--- the disk maximality --- in accordance with the surface brightness dependence
found by \citet{2013A&A...557A.131M}.  The inflection of the result seen at
$\qrw^{1.5h_R}>2$ is due to the stellar disk becoming less stable than the gas
disk.  For $\qrw^{1.5h_R}<2$, the predicted relation is very well described by a
power-law slope of -2.07 (dot-dashed line), which is steeper than the slope of
-1.54 predicted by \citet{2006ApJ...639..879L}.

For Figures \ref{fig:fig2}b and \ref{fig:fig2}c, we calculate ``effective'' mass
surface densities ($\Sge$ and $\Sse$) by integrating the stellar and gas mass
profiles from our dynamical model to $R_{25}$ and dividing by the total surface
area.  The Figures show the SFE ($\Ssfre/\Sge$) as a function of $\Sse$ and
$\qrwmin$.  The SFE is correlated with $\Sse$ and has a power-law dependence
that is in agreement with the empirical findings of \citet[][gray dashed line;
see their equation 6]{2011ApJ...733...87S} and the theoretical prediction of
\citet[][black dotted line; $\Ssfre/\Sge \propto \Sse^{1/2}$, where we have
optimized the normalizing constant]{2010ApJ...721..975O}.  Contrary to previous
results \citep{2008AJ....136.2782L}, we also find an albeit weak
anti-correlation between the SFE and the minimum disk stability.  This
anti-correlation is consistent with the expected {\it linear} relationship from
\citet[][dotted line with a best-fitting intercept]{2006ApJ...639..879L}, but
far from verifies it due to the scatter in the data.

\medskip \noindent{\bf Conclusion:}  Our analysis of the kinematic data from the
DiskMass Survey yields a significant anti-correlation between the star-formation
activity of a disk and its gravitational stability.  However, we also find that
our data are consistent with the equilibrium solution derived by
\citet{2010ApJ...721..975O}, with no significant correlation between
$\Ssfre/\Sge\Sse^{1/2}$ and any other quantity in our analysis.  In so far as
disk stability quantifies the susceptibility of a disk to modal potential
perturbations (bars, spiral arms, etc.), this result suggests that such
perturbations may not prohibit this proposed equilibrium.  We find an
error-weighted geometric mean of $\langle \log(\Ssfre/\Sge\Sse^{1/2}) \rangle =
-3.25\pm0.27$ in units of $(G/{\rm pc})^{1/2}$, where $G = 4.30\times10^{-3}$
(km/s)$^2$ pc $\msol^{-1}$ is the gravitational constant.  However, there is an
order-of-magnitude range in $\Ssfre/\Sge\Sse^{1/2}$ among the galaxies in our
sample.  It is of great interest to understand this scatter.

\acknowledgements Support for this work was provided by the National Science
Foundation (AST-0307417, AST-0607516, OISE-0754437, and AST-1009491), the
Netherlands Organisation for Scientific Research (614.000.807), NASA/{\it
Spitzer} grant GO-30894, the Netherlands Research School for Astronomy, and the
Leids Kerkhoven-Bosscha Fonds.

\bibliography{westfallk}

\end{document}